# Mendelian randomization in a multi-ancestry world: reflections and practical advice

**Authors and Affiliations:**

Amy M. Mason[1,2], Verena Zuber[3,4,5], Gibran Hemani[6,7], Elena Raffetti[1,2,8], Yu Xu[1,2,9,10], Amanda H.W Chong[6,7], Benjamin Woolf [6,11,12], Elias Allara[1,2,13], Dipender Gill[14], Opeyemi Soremekun[15,16], Stephen Burgess[1,12]

1. British Heart Foundation Cardiovascular Epidemiology Unit, Department of Public Health and Primary Care, University of Cambridge, UK
2. Victor Phillip Dahdaleh Heart and Lung Research Institute, University of Cambridge, UK
3. Department of Epidemiology and Biostatistics, School of Public Health, Imperial College London, London, UK
4. MRC Centre for Environment and Health, School of Public Health, Imperial College London, London, UK
5. UK Dementia Research Institute, Imperial College London, London, UK
6. Medical Research Council Integrative Epidemiology Unit, University of Bristol, Bristol, UK
7. Population Health Sciences, Bristol Medical School, University of Bristol, Bristol, UK
8. Department of Global Public Health, Karolinska Institutet, Sweden
9. Cambridge Baker Systems Genomics Initiative, Department of Public Health and Primary Care, University of Cambridge, Cambridge, UK
10. Institute of Precision Medicine, The First Affiliated Hospital, Sun Yat-sen University, Guangzhou, China
11. School of Psychological Science, University of Bristol, Bristol, UK.
12. MRC Biostatistics Unit, University of Cambridge, Cambridge, UK.
13. Health Data Research UK Cambridge, Wellcome Genome Campus and University of Cambridge, Cambridge, UK
14. Department of Epidemiology and Biostatistics, School of Public Health, Imperial College London, London, UK.
15. Department of Clinical and Biomedical Sciences, University of Exeter Medical School, RILD Building, Royal Devon & Exeter Hospital, Barrack Road, Exeter, EX2 5DW, UK
16. Molecular Bio-computation and Drug Design Laboratory, School of Health Sciences, University of KwaZulu-Natal, Westville Campus, Durban, South Africa.

**Corresponding Author:**
Dr Amy M. Mason
Address: Cardiovascular Epidemiology Unit, University of Cambridge, Victor Phillip Dahdaleh Heart and Lung Research Institute, Papworth Road, Cambridge CB2 0BB
Email: am2609@cam.ac.uk
Telephone: 07825299644
Orcid: 0000-0002-8019-0777

# Abstract

Many Mendelian randomization (MR) papers have been conducted only in people of European ancestry, limiting transportability of results to the global population. Expanding MR to diverse ancestry groups is essential to ensure equitable biomedical insights, yet presents analytical and conceptual challenges. This review examines the practical challenges of MR analyses beyond the European only context, including use of data from multi-ancestry, mismatched ancestry, and admixed populations. We explain how apparent heterogeneity in MR estimates between populations can arise from differences in genetic variant frequencies and correlation patterns, as well as from differences in the distribution of phenotypic variables, complicating the detection of true differences in the causal pathway.

We summarize published strategies for selecting genetic instruments and performing analyses when working with limited ancestry-specific data, discussing the assumptions needed in each case for incorporating external data from different ancestry populations. We conclude that differences in MR estimates by ancestry group should be interpreted cautiously, with consideration of how the identified differences may arise due to social and cultural factors. Corroborating evidence of a biological mechanism altering the causal pathway is needed to support a conclusion of differing causal pathways between ancestry groups.

# Keywords:

# Glossary

**Ancestry-defined population** – A large-scale grouping of individuals defined by shared patterns of genetic variation, typically identified through clustering analyses of genetic data. Ancestries are typically defined at a continental scale. Study populations drawn from the same ancestry population are expected to have broadly similar allele frequencies and patterns of linkage disequilibrium (LD).

**Linkage disequilibrium (LD)** - Patterns of correlations between genetic variants.

**Population structure** – The presence of distinct genetic subgroups within a population, demonstrated by differences in allele frequencies and linkage disequilibrium (LD) patterns. If ignored, population structure can lead to confounding known as population stratification, where genetic associations arise due to differences between subgroups.

**Admixed individual** – An individual whose genome reflects genetic contributions from multiple ancestry-defined populations. While all humans are admixed to some degree, the term typically refers to relatively recent mixing (within ~30 generations) between continental-scale ancestral groups.

**Ethnicity-defined population** – A self-identified cultural group. Unlike ancestry, which is ascertained by genetics, ethnicity is shaped by cultural and social factors. A population defined by ethnicity may contain individuals from multiple ancestry groups; similarly, a population defined by genetic ancestry may include individuals from multiple ethnicities.

**Geographically-defined population** – A group of individuals associated with a specific geographic location, such as through citizenship, place of residence, or country of birth. Such a population may or may not align with a specific ancestry or ethnicity group.

**Racially-defined population** - A socially constructed categorization of individuals, often informed by physical characteristics such as skin colour. Racial categories may be externally imposed and shaped by historical and political contexts, rather than arising from self-identification. Populations defined by race may or may not coincide with populations defined by ancestry, ethnicity, or geography.

**Target population** – The population of interest to the researcher – the group to whom the findings of a study are intended to apply.

# Introduction

Many published Mendelian randomization (MR) papers have the same limitation – the research was only conducted in people of European ancestry and so results may not be applicable to other ancestry groups. Despite common awareness of this limitation, the scientific community has struggled to develop approaches to unpick the life-long genetic effects of ancestry from the potentially modifiable social confounders of ethnicity (see **Glossary**). Our treatment of ancestry is in stark contrast to our treatment of sex. MR researchers are comfortable designing and interpreting both sex-combined and sex-stratified analyses[1-3] despite sex also being a complex genetic variable, with analyses potentially confounded by hormone mediated pathways impacting how our bodies function and a myriad of social factors varying by gender. However, we are less confident designing and interpreting MR analyses considering non-European ancestry populations.

We need to be precise and cautious in how we describe population groups. In a scientific context, ethnicity refers to a self-defined group, whereas ancestry refers to a researcher-defined group of people with similar genetic backgrounds[4] (see **Glossary**). In this paper we avoid the term 'race' – while it also reflects socially constructed groupings, race does not rely on self-identification and racial categories are not defined consistently across global contexts [4]. Additionally, categorizations by race perform poorly in terms of defining groups with similar genetic characteristics; genetic variation is greater within racial categories than between them[5,6]. In genetics research, people are clustered via genetic similarity thresholds set by researchers. These researcher-defined groups often exclude individuals who do not fit into a single cluster. Groups are typically given descriptive geographical or ethnic labels, although such groups are often not a representative sample of that population. Crucially these groups are not evidence of a biological reality to divisions by race or ethnicity. Many socially defined populations do not fit neatly into the conventional global reference ancestry clusters, such as the highly admixed population of Brazil [7].

In this work, we consider analyses using datasets from population groups with potential differences that complicate MR investigations. Differences may occur because of variation in genetic architecture, heterogeneity in underlying effects, or differences in the distribution of phenotypic variables[8] (**Figure 1** and **Table 1**). For example, there may be differences in genetic associations estimated in Bangladeshi populations living in Bangladesh versus diaspora Bangladeshi populations living in the UK, despite the groups having similar ancestry and ethnicity. The most appropriate target population depends on the research question and needs careful definition: geographic definitions may be preferred for environmental exposures, ethnicity definitions for cultural exposures, and ancestry-based definitions where the primary interest is genetic effects. We encourage accurate reporting of how populations are defined, particularly whether this is based on self-reported status or researcher-defined using genetic data. However, the discussions in this paper are relevant to all definitions.

The aim of an MR study is to identify genetic variants that mimic interventions on the distribution of a risk factor (the gene—environment equivalence assumption[9]), and to interpret associations of such variants with disease outcomes as informative of the potential effect of intervening on that risk factor in practice[10]. If genetic associations arise due to population stratification or confounding with nationality, then there is high potential for misleading results[11,12] Hence researchers are encouraged to use ancestry-specific datasets rather than multi-ancestry datasets. Two sample MR analyses typically rely on summary-level data from genome-wide association studies (GWASs) on the exposure and outcome from two samples that are ideally non-overlapping. They assume similarity between the GWAS populations for the exposure and outcome datasets[13]. Since the largest ancestry-specific genetic datasets are for European ancestry populations[14], analyses are often performed in European-only datasets. While this is a logical solution to the limitations of MR, this is criticized for its potential to deepen inequalities through optimizing health treatments for European patients.

This review discusses challenges conducting MR analyses that arise when using data from non-European populations, mismatched ancestry populations, multi-ancestry populations, and when combining estimates across different populations. We discuss motivations for performing analyses in different populations, illustrate how ancestry may impact MR analyses, and present practical analysis strategies when limited data are available.

# Why perform Mendelian randomization in diverse populations?

There are strong reasons to perform MR analyses in non-European ancestry groups. Firstly, to check for a non-null causal effect in a specific ancestry group, which may be challenging when data has a smaller sample size. Second to test whether a Mendelian randomization estimate is consistent across multiple ancestries, to support transportability of results[15] or identify potential heterogeneity in effects. If homogeneous, pooling estimates may allow greater precision of estimates than working in European data alone. This is complicated by heterogeneity in genetic architecture such as allele frequency distributions or linkage disequilibrium (LD) structure, and differences in environmental exposures. Finally, advanced analyses may allow researchers to investigate heterogeneity and identify potential effect modifiers between populations.

We expect differences in causal impacts of exposures due to underlying differences in biology between ancestry groups to be unusual[16]. Researchers need to consider the potential biases arising from ancestry and population structure which may lead to invalid inferences[17], as well as other population factors that would lead to differences between estimates. For an overview how these may impact MR analysis we refer to **Figure 1** and **Table 1,** with further details in the next sections.

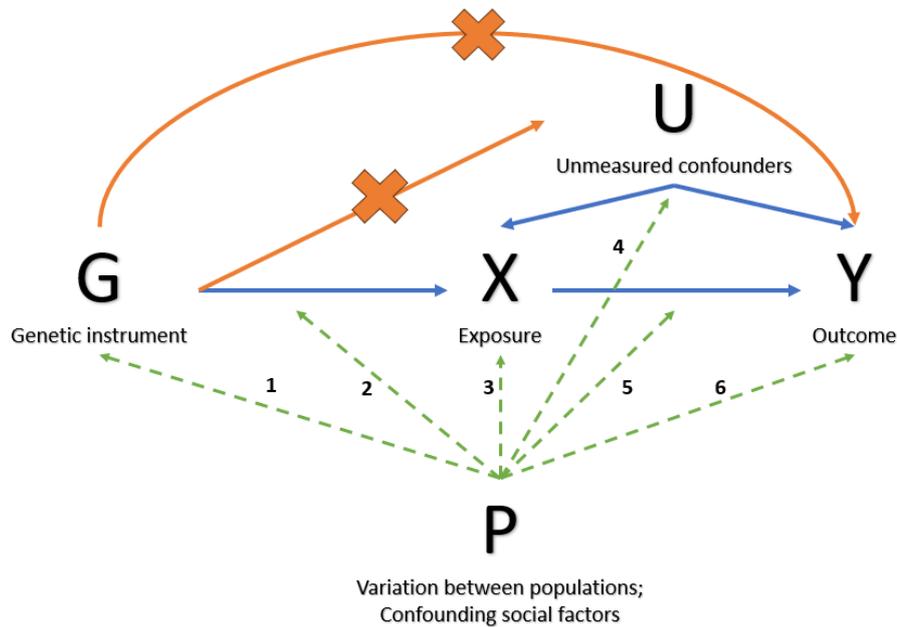

*Figure 1 Causal diagram of the MR assumptions and their possible violations. Blue arrows indicated the pathways expected in a MR analysis; orange arrows indicate potential violations of the exclusion restriction assumption. Green arrows (indexed by number) indicate potential sources of variation between populations that would lead to differences in genetic associations, and therefore differences in MR estimates and (potentially) inferences. Further details are given in Table 1.*

| Index in **Figure 1**. | Factors leading to potential differences in MR estimates between populations | Example |
|---|---|---|
| 1 | Differences in genetic architecture: allele frequency, linkage disequilibrium patterns (genetic correlations with nearby variants) | In African ancestry populations, there are associations between *MYH9* variants and chronic kidney disease risk, due to the high LD with *APOL1* variants.[18] In European ancestry populations, these variants are not in high LD with the *APOL1* variants. Misidentification of causal variants can lead to loss of power (with non-target selection data) or bias in MR estimates (with non-target exposure data)[19] |
| 2 | Heterogeneity in genetic effects | *CYP2A6* variants can be used as instruments for heavier smoking proxied by cotinine levels, but *UGT2B10* variants common in African ancestry populations and rare in Europeans alters cotinine metabolism independently of nicotine intake.[20] This interaction could give rise to apparent heterogeneous gene-exposure associations with smoking heaviness proxied by cotinine levels, even when the smoking levels are similar. This could give rise to heterogeneous MR estimates between ancestry groups even if the |

| | | impact of smoking on the outcome is homogeneous, making it difficult to assess whether the true causal effect varies between ancestry groups. |
|---|---|---|
| 3 | Differences in distribution of exposure | Genetic predictors of folate are associated with disease outcomes in populations with low folate diets but not in high folate ones [21]; thus MR analyses of impact of folate changes in high folate populations are unlikely to find evidence of effects even when MR analyses in low folate population do find effects. This can be attributed to a threshold effect of folate. |
| 4 | Differences in the distribution and effects of confounders | Variants in *FADS* genes influence the metabolism of PUFA (polyunsaturated fatty acids). These variants are associated with lower weight in Greenland populations who have high-PUFA diets, but not European populations who have low-PUFA diets [22], with evidence of an interaction between diet and genetics [23] Hence MR estimates using these variants as instruments for BMI on an outcome are likely to be heterogeneous for populations with different diets.<br><br>Differing environmental backgrounds may also make pleiotropic pathways harder to detect. Variants in *CHRNA5* are associated with alcohol intake and smoking heaviness [24]. If associations are evaluated in a non-smoking or low-smoking population, the impact on smoking might not be detected. |
| 5 | Differences in causal effect of exposure on outcome | The impact of alcohol on oesophageal squamous cell cancer risk is higher among people with variant rs671, which is common in East Asians, and uncommon in Europeans. This increase in impact on oesophageal cancer is not through increased alcohol consumption, but through changing how alcohol is metabolized. The variant causes acetaldehyde accumulation, magnifying the effect of alcohol per unit on the oesophagus, leading to different average causal effects of alcohol on oesophageal cancer risk in ancestry groups with differing distributions of rs671 [25-27] |
| 6 | Differences in definition of outcome | Diseases such as schistosomiasis have lower prevalence in European populations[28] due to geographic location. This means much lower power to detect putative causal effects on this outcome in some populations. Varying power may create an impression of heterogeneity in estimates as only some populations give convincing non-null results. |

*Table 1 Descriptions and examples of the potential sources of variability as indexed in Figure 1.*

## Differences in distribution of risk factors and confounders

Ethnicity is associated with strong disparities in many health outcomes. This is often for social and environmental reasons rather than genetic differences[29]. Certain traits (which may represent risk factors and/or confounders) vary in their distribution between populations. For example, alcohol consumption[30] and smoking behaviour[31] vary substantially across populations. These differences may change over time via societal factors such as public health policies[30], or arise due to environmental factors such as raised arsenic exposure in Bangladesh due to soil conditions[32]. They may remain localized to populations in specific geographic areas, or they may persist after migration. They may wane over time[33-35], further blurring the boundaries of what is attributable to ancestry, nationality, or ethnicity.

Differences in the exposure distribution could mean that MR estimates differ, particularly if there is a non-linear or threshold effect of the exposure on the outcome[36]. Gene-by-environment interactions (GxE) could lead to heterogeneity if the distribution of the environment variable differs between populations. Generating estimates solely in European ancestry populations who reside in Western countries may mean we miss crucial information about GxE interactions or deviations from linearity, leading to inaccurate assessment of causal effects in other populations. For example, genetic predictors of folate are associated with disease outcomes in populations with low folate diets but not in high folate ones, suggesting a threshold in the effect of folate supplementation[21] that is missed if it is assessed solely in a Western population. Such differences may also inform us about the nature of the causal effect; for example, a threshold effect of folate supplementation revealed in analyses of low-folate populations would be relevant to individuals in Western countries with low folate levels.

## Differences in distribution of outcomes

Prevalence of the outcome of interest in a population may impact the power to estimate causal effects. Some diseases, such as schistosomiasis[28] and systemic lupus erythematosus[37] are rarer in European ancestry populations and thus our power to detect causal effects is reduced when using data from these populations. Bias in diagnosis cut-offs can also cause differing ancestry prevalence; HbA1c measurements are lower in people with sickle-cell trait, leading to underdiagnosis of diabetes in African-ancestry populations.[38]

Differences in outcome prevalence may also impact clinical advice. Anticoagulants are known to reduce risk of stroke in European ancestry groups by preventing blood clots.[39] However, East Asian ancestry datasets show higher rates of intracerebral haemorrhage[40,41], and higher risks of bleeding with anticoagulant use[42], which can diminish the overall population-level benefit.[39] It is unclear to what extent the difference in stroke rates is driven by differences in genetic burden or societal factors, and thus the optimal treatment for individuals of East Asian ancestry living in Western countries is unclear. One of the benefits of using MR in different populations would be the potential to unpick the reasons for such differences.

## Heterogeneity in genetic associations

The transportability of GWAS estimates across populations varies between traits. For example, of 802 known European height variants, 643 (80%) showed consistent direction and 205 (26%) were nominally significant (p<0.05) in a ~60k African ancestry population[43] while 1,177 of 3,806 known height variants (31%) and 336 of 1,247 known BMI loci (26.9%) were similarly consistent in direction and nominally significant in a ~60k Hispanic/Latino population [44] This seems low, but part of this is due to lower power in the non-European populations, rather than true differences in estimates. When comparing South Asian and European ancestry populations, Huang et al found that they observed similar numbers of significant variants to what they expected, once they accounted for the relative power of the GWAS with different size datasets.[45] Comparison of European and African ancestry populations in the UK found little evidence of gene-environment or gene-gene interactions across 47 phenotypes, suggesting that mutations generally operate similarly in different ancestry populations; interactions may cause variation in effect size but rarely in a way specific to ancestry groups within the same geographic area.[46]

As an example of ancestry specific heterogeneity, using cotinine levels as a proxy biomarker for heavy smoking is complicated by *UGT2B10* variants common in African ancestry and rare in Europeans. These alter cotinine metabolism independently of smoking levels[20], creating heterogeneous gene-exposure associations. Where heterogeneous estimates exist, they are a reason to be cautious in using multi-ancestry GWAS data to calculate MR estimates, as the exposure-association estimate is dependent on the ancestry composition of the dataset.

## Differences in genetic architecture

Different ancestry groups may have different genetic architecture; differing LD structures and allele frequencies[47], with common variants from one population extremely rare or absent in other populations. Humans are 99.9% genetically identical with only 0.1% of our DNA base pairs differing from person to person[47]. If we divide people into ancestry groups, more of the variation is found within an ancestry group than between them - estimates suggest only 5-23% of the variance is between ancestry groups [48,49]. We expect differences in causal impacts of exposures due to underlying differences in biology between ancestry groups to be unusual[16], but exposure levels will vary by ancestry groups due to differences in allele frequencies, although this variation is usually much smaller than that from environmental and cultural factors.. For example, South Asian ancestry populations have a higher genetic burden (higher frequencies of risk alleles) for beta cell dysfunction related predictors of diabetes compared to European ancestry populations[50,51]. This may explain the higher prevalence of diabetes in younger, low BMI South Asian people[51] and need for different treatment approaches[52] without necessarily implying that the causal impact of BMI or beta-cell dysfunction on diabetes differs between ancestry groups.

MR is complicated by differing genetic architecture. If an instrument used in MR is not the causal variant, but rather is correlated with the causal variant, then this correlation may be stronger or weaker in different ancestry groups where LD-patterns differ, leading to bias in two-sample MR analyses [19].

Differences in genetic architecture also provide opportunities. There are common population-specific functionally-relevant genetic variants with large effects influencing Lp-PLA$_2$ in the *PLA2G7* gene for South[53] and East Asians[54], triglycerides in the *APOC3* gene in Amish populations[55], and tolerance of alcohol in East Asians[56]. These variants allow powerful MR analyses in these populations that cannot be performed in conventional European-only analyses.

# Implementing Mendelian randomization in diverse populations

## Assumptions required

We primarily consider assumptions required for a MR estimate to be a reliable guide as to the presence and direction of a causal effect, rather than those which predominately affect the size of the causal estimate. While MR does provide a causal estimate, it typically relates to a small and lifelong difference in the distribution of the exposure, whereas any intervention performed will typically be larger, limited in duration and performed on a mature target population[57]. Hence MR estimates are more suited to providing evidence supporting or refuting the existence of a causal effect, rather than estimating the potential impact of an intervention in practice.

Standard MR investigations require three datasets, which are ideally sampled from the same homogeneous population, to: 1) select genetic variants, 2) estimate associations with the exposure, and 3) estimate associations with the outcome. These datasets may be completely distinct (three-sample MR), they may be the same dataset (one-sample MR), or they may partially overlap[58]. For valid inferences, each genetic variant must satisfy the assumptions of an instrumental variable in the outcome dataset: relevance (the variant must be associated with the exposure), exchangeability (no confounding pathway between the variant and outcome), and exclusion restriction (any causal pathway from the variant to the outcome passes via the exposure)[59]. If a genetic variant is not associated with the exposure in the target population, then the variant is not a relevant instrument for that population. When an MR analysis is based on multiple genetic variants, irrelevant instruments may not invalidate inferences[60], but will add noise to the analysis. Similarly, genetic associations with the exposure varying in magnitude between datasets leads to inefficiency, but not invalidity[61].

More concerning is the possibility that a genetic variant violates the instrumental variable assumptions in one population group. One such reason for violation in one population group but not another is linkage disequilibrium with a pleiotropic variant. Another potential reason is differences in the distribution of risk factors, which lead to associations with a confounder in one dataset that are not present (or not detected) in the other dataset.

Hence, from the perspective of validity, the ideal MR analysis for a specific population would select variants and estimate associations with the exposure and outcome in datasets representing the target population.

## Little or no available data of the target population

However, suitable datasets representing the target population may not exist. If they do exist, the best variants to use as instruments may not be available or be less accurately imputed - genetic tools are often designed using European ancestry populations, and their adaption to other populations can be slow[62]. Missing variants can be proxied by highly correlated variants, if the data for population-specific LD-patterns are known.

There may not be multiple non-overlapping datasets available to select variants and separately estimate genetic associations. Sample overlap can lead to an overestimation of the genetic association due to random sample variation (winner's curse bias), which is particularly concerning when there is overlap between data used for selection of variants and estimation of genetic associations with the outcome[63]. Overlap between the datasets used for estimating genetic associations with the exposure and outcome means that weak instrument bias is in the direction of the confounded association between the exposure and outcome, potentially leading to false positive findings[64]. If researchers have access to individual-level data, these concerns can be mitigated by the use of cross-validation methods[65,66]. Otherwise, a compromise may be required between performing the analysis using non-overlapping datasets to reduce bias and using all available data to improve power.

The best approach to take will depend on the specifics of the analysis and the available datasets. If multiple large non-overlapping datasets are available with sufficient power to detect a clinically relevant causal effect, then they might be preferred even if this sacrifices some statistical power. On the other hand, if restricting the analysis to non-overlapping datasets would lead to inadequate power, then it may be better to use overlapping datasets at the risk of some bias.

As well as being less numerous, genetic datasets for non-European ancestry groups may also have smaller sample sizes. Many national biobanks, such as the Ugandan Genome Resource[67] (N=6,500) and the Mexican Biobank[68] (N=6,000), are significantly smaller than UK Biobank (N=500,000) . Small sample sizes have a double impact on the power of MR analyses: first, it reduces the number of genetic variants associated with the exposure and hence the proportion of explained variance in the exposure[69], and second, it reduces the precision of genetic associations with the outcome[70]. While considerable effort and huge investment is going into creating larger, more diverse biobanks, this is only a partial solution. For some populations, the total number of individuals is simply too small to allow for sufficiently powered recruitment. For example, the entire living population of Iceland is smaller than UK Biobank[71] and so similarly precise estimates for the Icelandic population as for a wider European ancestry population cannot be obtained. Insufficient data is likely to always be a problem for some exposures and population groups, particularly the most expensive and difficult-to-obtain measurements, such as tissue-specific protein levels.

Equally, even if distinct datasets are available for a target population group, we may want to leverage data from non-target population groups to improve the precision of estimates and the efficiency of the analysis.[72] A balance is often required between restricting the analysis to the most relevant data and ignoring potentially relevant data.

# Practical analysis strategies

## Describe data clearly

In choosing datasets, researchers should consider how the population under analysis has been defined, and how well that agrees with the target population for their research. Researchers need to understand the characteristics of each dataset they use, how data were collected, and the dataset's sociodemographic composition. There could be factors other than ancestry that vary between different populations – for example, average age[73], year of birth, gender balance, affluence, and rural vs urban populations. Outcome definitions and screening strategies may also vary across studies or populations. Because of this, we need to be cautious of attributing differences between results to ancestry differences. Hou et al found very high correlation of genetic associations across local admixed ancestry populations compared to trans-biobank ancestry populations, suggesting that large differences may be due to environmental differences.[74]

We proceed to describe different approaches to the selection of datasets for the implementation of MR, with an overview in **Figure 2.**

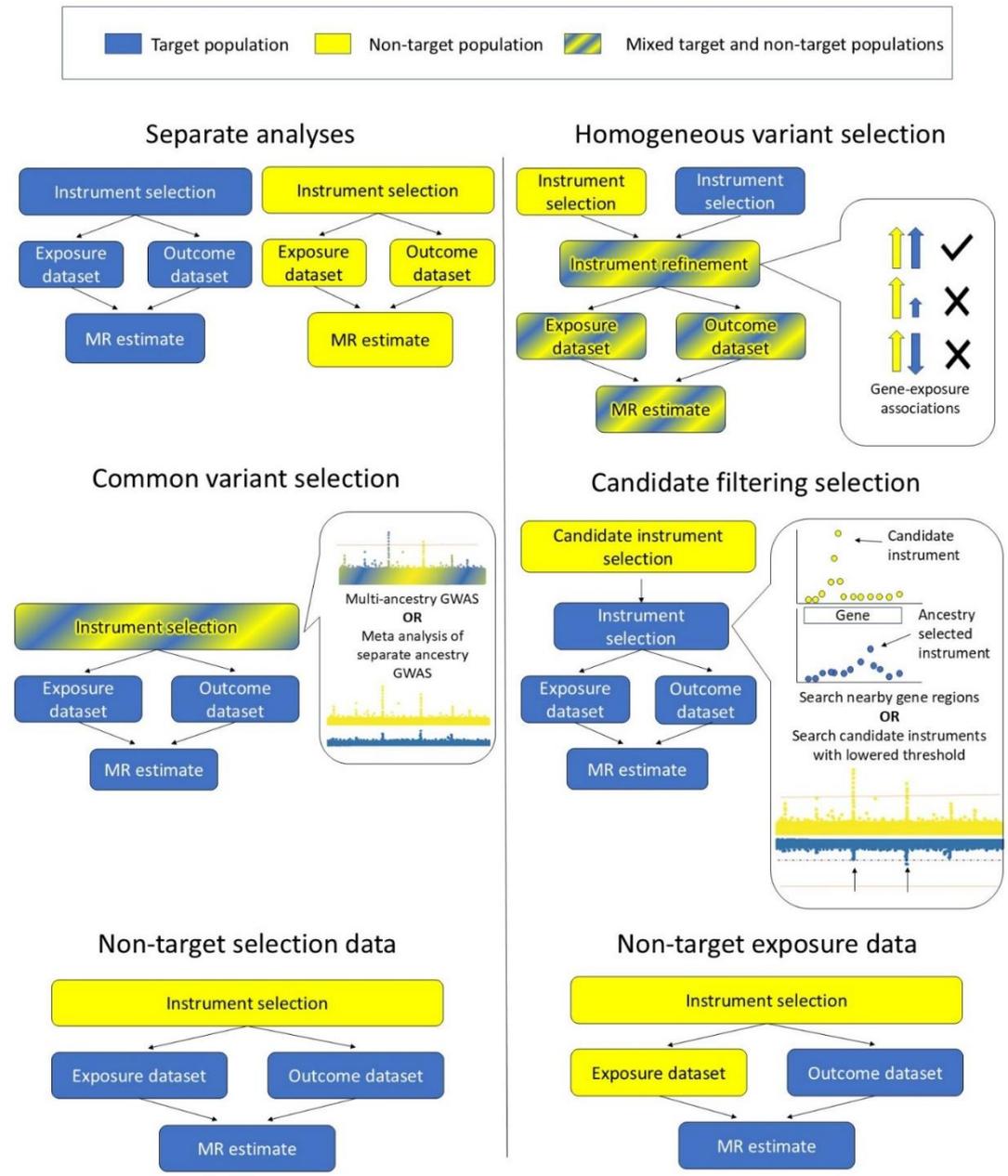

*Figure 2  Possible strategies for selecting genetic instruments and conducting Mendelian randomization (MR) analyses in a multi-ancestry context. Colour of boxes indicate the source population for instrument selection, exposure, and outcome data: blue for the target population (usually non-European ancestry), yellow for a non-target population (usually European ancestry), and striped for a mixed-ancestry population that includes the target population. Further discussion of the limitations and benefits of these approaches is provided in the text.*

## Separate analysis in each target population

The simplest approach is to perform all analyses within datasets of the target population, with instrument selection, and genetic associations with exposure and outcome all coming from suitable homogeneous datasets. For example, Zheng et al.[75] ran a parallel analysis on risk factors for chronic kidney disease performing one analysis using data exclusively from East Asian ancestry and another using data solely from European ancestry populations. While this ensures any genetic variants used in an ancestry-specific analysis are relevant to the exposure in that population, the disadvantage of the approach is lack of suitable data creating difficulty in identifying genetic instruments for each ancestry group. Zheng et al.[75] could only examine 17 out of 45 risk factors of interest in the East Asian arm due to either lack of suitable GWAS data on the risk factor or a lack of suitable instruments in the existing GWAS.

## Selection for homogeneous variants, with combined analysis

With sufficient data from multiple populations, it is possible to select instruments with homogeneous associations across datasets. When such instruments can be found, there is potential for using aggregated multi-ancestry GWAS data for the exposure and outcome. This is potentially advantageous as it can include admixed individuals. However, a key limitation is the reduced power due to only including variants with consistent associations.

For example, Hamilton et al[76] selected rs2228145 as an instrument for C-reactive protein from a European ancestry study, then assessed the variant for consistency as a trans-ancestry instrument. MR estimates were obtained using multi-ancestry GWAS for both the exposure and outcome datasets.

## Common variant selection

This approach uses an aggregated dataset containing members of the target population(s) for the selection of variants. We then apply this instrument to perform separate MR analyses using genetic association estimates obtained in each target population. The instrument selection data may be a single multi-ancestry GWAS or a meta-analysis of several single-ancestry GWAS via methods such as TEMR[77] and CAMERA[19]; . This approach assumes the same genetic variants associate with the exposure in each population but permits variation in the size of the association.

The main advantage is that the genetic variants are chosen based on a larger sample size, and hence the analysis potentially has greater power. Another benefit where several populations are being compared is that the instrument choice remains consistent between population-specific analyses, meaning that differences between estimates arise only due to different genetic associations in the datasets. The main disadvantage is, as the instrument is not validated specifically in the target population, irrelevant variants and variants with very low frequency in the target population may be retained. If the analysis is based on variants in the same gene region, there may be correlation between the variants. This can be managed with correlated-instrument methods[78], with the caution that the correlation structure may differ between populations. Another disadvantage is that if the mixed population data is dominated by a non-target population group, variants with associations specific to the target group may be missed.

Morris et al.[79] selected variants from a combination of large GWAS datasets representing different ancestry populations, to create a single instrument set. They ran analyses separately using exposure data from three different ancestry populations, and the same outcome data (primarily from European ancestry populations). However, given that the critical aspect of a MR investigation is assessing genetic associations with the outcome, this represents three differently weighted analyses for the same outcome population (i.e. Europeans), rather than distinct ancestry-specific analyses.

## Candidate filtering selection

A variation on this is to use biological information, a large non-target population GWAS, or aggregated GWAS data to identify gene regions containing genome-wide significant genetic predictors of the exposure, then to filter down to target population-specific variants an analysis of the target population(s). For example, the non-target population dataset may identify variants that are low frequency in the target population, but nearby variants in the same gene region may also associate with the exposure and appear more frequently in the target population. The key advantage over a separate analysis is a less strict significance threshold can be used for validation in the target population, mitigating somewhat the problem of smaller sample sizes. Irrelevant and population-specific correlated instruments can also be excluded.

A simpler version of this approach is to filter the list of the genome-wide significant genetic variants from a non-target population GWAS by validating the candidate variants in the target population. This still justifies using a less strict significance threshold but may miss potentially more suitable instruments in the wider regions.

For example, Hamilton et al[80] identified variants in a large multi-ancestry meta-analysis, then used ancestry specific GWAS and ancestry specific LD data to retain only independent variants with the same direction of effect as the multi-ancestry GWAS, even if they did not achieve GWAS significance in the ancestry specific GWAS. These instruments were applied in MR analyses with ancestry matched exposure and outcome data and the results meta-analysed.

## Selection based on non-target population

An alternative is to base variant selection on a dataset from a non-target population, but to use associations with the exposure and outcome in the target population. This may be desirable if very limited exposure data exists for the target population. The larger non-target dataset will likely result in more variants being selected but may also result in the selection of irrelevant or pleiotropic instruments. Where possible, population-specific pleiotropy should be considered. This is not only because there may be novel biological pathways unique to certain populations, but because gene-environment interactions may conceal pathways in some populations. For example, the variant rs1695 is associated with asthma but its effect may be mediated by pollution levels[81], and thus may not be a relevant instrument in low pollution populations.

Park et al. [82] used an instrument derived in a "White British" ancestry population, and then applied it using two multi-ancestry datasets for the exposure and outcome associations. They checked the genetic variants remained predictive of the exposure in the multi-ancestry datasets but did not consider ancestry-specific LD or allele frequencies.

## Using non-target exposure data

Where data in the target population is extremely limited, it may not be possible to find suitable exposure-association data. Using genetic association data for the exposure that comes from a population that differs from the target population assumes that the existence and direction of genetic associations[83], and the LD patterns between the causal variants and the instruments[19,72] are consistent between populations. If this holds, MR with non-target exposure data will be a valid assessment of the effect of the genetically predicted exposure on the outcome in the target population. [76,77], with any population-mismatch bias typically towards the null and proportional to the genetic distance between the populations[72], Instrument consistency can be tested e.g. MRSamePopTest[13], CAMERA[84], XMAP[85] and similar.

Provided valid instruments are employed, it is the ancestry group of the genetic associations with the outcome that determines the interpretation of the analysis. This is because a Mendelian randomization analysis tests whether variants are associated with the outcome. The exposure-associations change the relative weightings of the variants and may alter the magnitude of the estimate, but only the outcome-associations determine whether there is evidence for a causal effect of the exposure. Hence, if an analysis uses non-target outcome data, it would not be reasonable to describe this as a MR analysis in the target population.

## Interpretation of results

Researchers should be cautious in using MR alone to justify claiming a difference in causal effects between populations. Evidence should be triangulated with lab-based studies, observational data, and randomized controlled trials. Consideration should be given to potential confounding by environment, behaviour, or deprivation before attributing differences to ancestry (see Figure 1). Data from diaspora groups may be useful in unpicking whether heterogeneity in estimates is attributable to geographic factors.

Researchers should be clear as to the limits of the applicability of their results, and should remain alert for how titles, abstracts and graphs may be misinterpreted in absence of wider context. Poorly worded labels and titles may cause confusion for clinicians and policy makers. There is a temptation to conflate ethnicity and ancestry labels, risking application of results to individuals who are outside the population under analysis; this approach may cause more harm than universally applying estimates derived from large European ancestry populations. If we claim that an effect has been demonstrated in South Asians, do we mean individuals living in South Asia, individuals who engage in South Asian cultural practices, or individuals with South Asian ancestry?

With suitable additional data, it may be possible to use advanced MR techniques to further investigate differing causal estimates between populations. MR-GxE[86] could be used as a sensitivity analysis to investigate other potential non-ancestry factors, such as deprivation[87] or nationality. Where populations differ significantly with regards to known confounders of the outcome or mediators of the causal effect, multivariable MR[88] could help unpick the true causal determinants of an outcome.

# Conclusion

All epidemiological analyses combine data on individuals that are in some way heterogeneous. Contextual judgement is required to decide whether it is preferred to analyse groups together or separately, whether those groups are defined according to age, sex, or ancestry. However, there are specific considerations due to population stratification that mean we should be cautious about blindly combining individuals with different ancestries in an MR analysis.

When it is plausible to assume that features may be homogeneous across populations, it may be reasonable to borrow data from other ancestry groups or attempt a combined analysis, while being clear on the assumptions made. This is particularly true for variant selection, as most genetic associations are concordant across population groups[46]. However, it is recommended to assess the relevance (that is, if associated with the exposure) and validity (that is, if not associated with the outcome via a confounding or pleiotropic pathway) of candidate instruments in the population under investigation. Keeping aspects of the analysis similar across population groups aids comparison of estimates.

We presented a range of strategies for combining or separating MR analyses dependant on the features of the instrument and populations under consideration. They have the limitation of relying on a discrete concept of ancestry and the availability of ancestry-specific GWAS data, with only one approach supporting the inclusion of admixed individuals. Development of new methods may be needed to manage data derived using continuous or local ancestry approaches.[89 90]

Researchers looking at differences in MR estimates by ancestry group should carefully consider what evidence they have of a plausible biological mechanism altering the causal pathway. They should remain alert to the possibility that any identified differences in estimates may arise due to social and cultural factors rather than differences in biological mechanisms. Evidence beyond MR is essential before concluding that causal effects vary between ancestry populations.

# Funding and Disclosures

**Sources of Funding:**

VZ gratefully acknowledges the United Kingdom Research and Innovation Medical Research Council grants MR/W029790/1. GH and AHWC work within the MRC Integrative Epidemiology Unit at the University of Bristol, which is supported by the UK Medical Research Council (MC_UU_00032/1). ER was supported by The Swedish Research Council for Health, Working Life and Welfare (Forte 2022-00882) and The Swedish Research Council (VR 2023- 01982). YX was supported by the UK Economic and Social Research Council (ES/T013192/1). SB is supported by the Wellcome Trust (225790/Z/22/Z) and the United Kingdom Research and Innovation Medical Research Council (MC_UU_00040/01).
This work was supported by core funding from the British Heart Foundation (RG/F/23/110103), NIHR Cambridge Biomedical Research Centre (NIHR203312) [*], BHF Chair Award (CH/12/2/29428), and by Health Data Research UK, which is funded by the UK Medical Research Council, Engineering and Physical Sciences Research Council, Economic and Social Research Council, Department of Health and Social Care (England), Chief Scientist Office of the Scottish Government Health and Social Care Directorates, Health and Social Care Research and Development Division (Welsh Government), Public Health Agency (Northern Ireland), British Heart Foundation and the Wellcome Trust.

 *The views expressed are those of the authors and not necessarily those of the NIHR or the Department of Health and Social Care.

**Conflicts of Interest**

DG and SB are employed by Sequoia Genetics, a private limited company that works with investors, pharma, biotech, and academia by performing research that leverages genetic data to help inform drug discovery and development.

No other authors have declared conflicts of interest.

**Data and code availability:** This is a review article and generated no new data or code. All data underlying this review are cited in the references.